\documentstyle[12pt,aasms4]{article}
\doublespace
\input epsf

\font\petit=cmr8 scaled \magstep0
\def\msol{M_\odot}

\def\mv{M_{\rm V}}
\def\mr{M_{\rm R}}
\def\mi{M_{\rm I}}
\def\mj{M_{\rm J}}

\def\mk{M_{\rm K}}
\def\te{T_{\rm eff}}
\def\lsol{L_\odot}
\def\simgr{\,\hbox{\hbox{$ > $}\kern -0.8em \lower 1.0ex\hbox{$\sim$}}\,}
\def\simle{\,\hbox{\hbox{$ < $}\kern -0.8em \lower 1.0ex\hbox{$\sim$}}\,}
\def\beq{\begin{equation}}
\def\eeq{\end{equation}}

\def\aj{AJ}                  
\def\araa{ARA\&A}             
\def\apj{ApJ}                 
\def\apjs{ApJS}               
\def\aap{A\&A}                
\def\aaps{A\&AS}             
\def\pasp{PASP}               

\begin{document}

\title{{\bf Evolutionary models for very-low-mass stars and
brown dwarfs with dusty atmospheres} }

\author{ {\sc G. Chabrier$^{1,2}$, I. Baraffe$^{1,2}$, 
F. Allard$^{1}$ \and P. Hauschildt$^{3}$ }}

$^{1}$ {Ecole Normale Sup\'erieure de Lyon, C.R.A.L (UMR 5574 CNRS),\\
\indent 69364 Lyon
 Cedex 07, France\\ \indent  email: chabrier, ibaraffe, fallard @ens-lyon.fr

$^{2}$  Astronomy department, University of California, Berkeley, CA 94720

$^{3}$Dept. of Physics and Astronomy and Center for Simulational Physics,
\\  \indent University of Georgia Athens, GA 30602-2451 \\  \indent email:
yeti@hobbes.physast.uga.edu}

\begin{center}
$\rm\underline{Submitted\ to}$: {\sl Astrophysical Journal}

\bigskip
$\underline{Version}$: \today
\end{center}

\date{Received /Accepted}

\bigskip
\bigskip

\begin{abstract}

We present evolutionary calculations for very-low-mass stars and brown dwarfs based on synthetic spectra and non-grey
atmosphere models which include dust formation and opacity, i.e. objects with 
$\te\simle 2800$ K. 
The interior of the most massive brown dwarfs is shown to develop a conductive core after $\sim 2$ Gyr which slows down their cooling.
Comparison is made in
optical and infrared color-magnitude diagrams with recent late-M and L-dwarf observations. The saturation in optical colors and the very red near-infrared colors of these objects are well
explained by the onset of dust formation in the atmosphere. Comparison of the
faintest presently observed L-dwarfs with these dusty evolutionary models
suggests that dynamical processes such as
turbulent diffusion and gravitational settling are taking place near the photosphere. As the effective temperature decreases below $\te\approx 1300-1400$ K,
the colors of these objects move to very blue near-infrared colors, a consequence
of the ongoing methane absorption in the infrared. We suggest the possibility of
a brown dwarf dearth in $J,H,K$ color-magnitude diagrams around this temperature.

\bigskip

\end{abstract}

\section{Introduction}

Since the unambiguous identification of the first cool brown dwarf (BD) Gl 229B 
(Oppenheimer et al. 1995), the discovery of objects with temperatures cooler
than the latest known M-dwarfs has been steadily increasing.  
These discoveries revealed two classes of objects requiring new
spectral classifications. On one hand, a family of objects shows
very red infrared ($J$-$K$, $H$-$K$) colors while the signature of metal oxides
(TiO, VO),
whose band strength index is used to classify M-dwarf spectral types, and hydrides (FeH, CaH bands) disappear gradually from the spectral distribution, as observed e.g. in GD 165B (Kirkpatrick et al. 1999a). Several of these
so-called  ``L'' dwarfs (Mart\'\i n et al. 1999; Kirkpatrick et al. 1999b),
 have been discovered by the
near-IR surveys DENIS (Delfosse et al. 1999) and 2MASS (Kirkpatrick et al. 1999b). 
On the other hand, a dozen of objects with properties similar to Gl229B
have been recently discovered with 2MASS (Burgasser et al. 1999), 
SDSS (Strauss et al. 1999) and the VLT (Cuby et al. 1999). The IR spectrum
of these objects is
characterized by the unambiguous signature of methane absorption in the H, K and L bands, the predicted dominant equilibrium form of carbon below a local temperature $T\approx 1300-1500$ K in the $P\approx 5-10$ bar range (Allard \& Hauschildt 1995; Tsuji et al. 1996; Fegley \& Lodders 1996). This yields blue near-infrared colors, with $J-K\simle 0$ but $I-J\simgr 5$ (see e.g. Allard 1999). These cool
dwarfs are identified as ``methane'' dwarfs, sometimes called also "T-dwarfs"
(Kirkpatrick et al. 1999b).

The spectroscopic and photometric properties of
L-dwarfs and, to less extent, of methane-dwarfs cannot be properly reproduced by dust-free atmosphere
models,
indicating that below $\te \, \simle 2800 K$, theoretical models must include
the formation and the opacity of dust grains (Tsuji et al. 1996, 1999; Ruiz et al. 1997;
Jones and Tsuji 1997; Allard 1998; 
Leggett et al. 1998; Tinney et al. 1998;
Basri et al. 1999; Kirkpatrick et al. 1999b).

In previous papers, we have developed a consistent theory of the
structure and the evolution of low-mass, dense objects
(Chabrier and Baraffe 1997, CB97) based on non-grey dust-free
atmosphere models (Hauschildt et al. 1999, HAB99). Evolutionary models
derived from this general theory
describe successfully various observed properties
of M-dwarfs down to the bottom of the main sequence: mass-magnitude relationships,
color-magnitude diagrams, mass-spectral type relationships (Chabrier et al. 1996; Baraffe \& Chabrier, 1996; Baraffe et al. 1997, 1998, BCAH98).
The aim of the present paper is to extend these calculations into the brown dwarf domain by including
the effect of grain formation both in the atmospheric equation of state (EOS) and in the opacity for objects with $\te \, \simle 2800 K$.
The first comprehensive description of the effect of grain formation on the evolution of BDs was done by Lunine et al. (1989) and has been updated and incorporated in Burrows et al. (1993, 1997) calculations. However only a few grains were included in these calculations and the grain opacity is treated either as a frequency-independent Rosseland mean,  
implying
grey atmosphere conditions, above 1300 K (see Figure 1 of Burrows et al. 1997),
or is simply ignored (assuming grains settle immediately below the photosphere) below 1300 K, where non-grey atmosphere models are used. The present calculations
extend significantly the number of grain species and include consistently their
frequency-dependent opacity in the transfer equation, yielding consistent non-grey atmosphere structure, spectral colors and evolutionary calculations.
The models are described in \S2, whereas comparison between theory and observation in various color-magnitude diagrams (CMD) is presented in \S3.
The remaining uncertainties and shortcomings in the theory are discussed in the conclusion.  

\section {Model description}

The main input physics involved in the present
calculations has been presented in details
in CB97 and BCAH98.
The outer boundary conditions between the interior and
the non-grey atmosphere profiles are described in CB97. Non-grey atmospheres are a necessary condition when absorption coefficients strongly depend upon frequency, which is the case for all objects below $\te \sim 5000$ K, the onset of molecular formation.
The formation of grains in the present model atmospheres is described briefly in
Leggett et al. (1998) and Allard (1999) and in details in Allard et al. (2000a). The equilibrium abundances of each grain species are determined from the Gibbs energies of formation, following the method developed
by Grossman (1972). At each temperature and for each
condensed phase under consideration, the variable $K_c(T)$ is calculated from the partial pressures of the species $i$ forming the condensate (e.g. $Al$ and $O$ for corundum $Al_2O_3$), determined by the vapor phase equilibria,
$P_i=N_i\mathcal{R}T$, where $N_i$ is the number of moles of species $i$ and $\mathcal{R}$ is the gas constant. This variable $K_c$
is compared to the equilibrium constant $K_{eq}$, calculated from the Gibbs energy of formation of the condensate. The abundance of a condensed species is obtained by the condition that this species be in equilibrium with the surrounding gas phase, $K_c\ge K_{eq}$ (Grossman, 1972). The opacities of the grains are calculated from Mie theory (see also Alexander \& Ferguson, 1994).

The formation of condensed species depletes the gas phase of a number of molecular species (e.g. VO, TiO, FeH, CaH, MgH) and of refractory elements such as Al, Ca, Ti, Fe, V (Fegley \& Lodders,
1996; Allard 1998; Lodders 1999; Burrows \& Sharp, 1999), modifying significantly the atmospheric structure and the emergent spectrum.
For late M-dwarfs and for massive and/or young BDs, the main cloud formation is
predicted to occur near the photosphere (see e.g.  Allard 1998; Lodders 1999).
The present
calculations assume a grain-size distribution in the submicron range (Allard et
al 2000a). This general grain treatment is consistent with the observations of the objects near the bottom of and below the
MS, namely the DENIS and 2MASS objects (Delfosse et al. 1999; Kirkpatrick et al. 1999b), GD 165B (Kirkpatrick et al. 1999a) and Kelu-1 (Ruiz et al. 1997), which all exhibit strong thermal heating
and very red colors. Asides from this complex thermochemistry, grain formation is
in fact a dynamical process and involves a balance between various timescales, such as the condensation, evaporation, coagulation, coalescence and convection timescales (see e.g. Rossow 1978), not mentioning the fact that non-equilibrium species might be present, as for example in the atmosphere of Jupiter (Fegley \& Lodders 1994). Since such dynamical processes are not incorporated in the theory at the
present stage, we elected to conduct calculations under various extreme assumptions.

The first set of models are based on the dust-free ``NextGen'' model atmospheres (Allard et al. 1996; HAB99) and are the same as in BCAH98. In the second set of models, {\it all} condensed equilibrium species are included both in the EOS and in the opacity,
taking into account dust scattering and absorption in the radiative transfer equation.
These models
are referred to as ``DUSTY'' models. The third set of models includes the formation
of grains in the atmosphere EOS, thus taking into account the photospheric depletion of dust-forming elements in the gas phase, but not in the transfer
equation, therefore ignoring the opacity of these condensates. This case, 
referred to as ``COND'' models, mimics a rapid gravitational settling of all grains below the photosphere. This is similar to the Burrows et al. (1997) 
calculations below $\te \le 1300$ K (see above).
 
As mentioned above, the
DUSTY and COND models represent extreme situations which bracket the more
likely intermediate case resulting from 
complex, and presently not understood, thermochemical and dynamical processes.

\subsection{Atmosphere profiles}

Figure \ref{fig1}
 displays $P$-$T$ atmosphere profiles for $\te$ = 1800 K,
log $g$=5 and solar composition, under the afore-mentioned different assumptions for the dust treatment. The atmospheric heating due to the large grain opacity (the so-called greenhouse or backwarming effect) in the DUSTY
model is clearly visible, and yields a significantly hotter structure. The effect is dominant 
in the outer layers, where the spectrum forms, and decreases
in the inner layers,
where the connection with the interior profile occurs (see also Tsuji
et al. 1996).
When dust
opacity is not taken into account as in the COND models, the inner atmospheric
structure is barely affected by the formation of dust and is similar
to the dust-free case (NextGen) for this temperature. Only for significantly cooler $\te$ does
dust formation start to affect substantially the inner structure compared
to dust-free models.

As indicated in Figure \ref{fig1}, convection occurs only in optically-thick regions ($\tau > 1$), in contrast to hotter objects
(see BCAH98). However, this optical depth corresponds to only about one pressure scale height $H_{\rm P}$, or even less, from the photosphere ($\tau \sim 1$)
near which occurs the condensation temperature of most grains for objects near the bottom of the MS.
 Even for the cooler models ($\te \sim 1000$ K) 
where convection retreats to deeper layers, the top of the convective zone lies only at a few $H_{\rm P}$ 
 from the photosphere. This can have important consequences on the formation and settling of atmospheric grains. Indeed, turbulent diffusion, as produced for example by overshooting from the convection zone or by advection due to rotation, is a much more efficient mechanism than microscopic diffusion and
sedimentation,
at least for submicron-size particles.
Although the temperature at the top of the convective zone is found to be
generally above the condensation temperature
of all grains, this turbulent diffusion 
will efficiently bring material upward to the region of condensation and maintain small-grain layers, which otherwise would have settled gravitationally, in this region. Note that for submicron-size species, the condensation time $\tau_{cond}$ is much smaller than other dynamical timescales (see e.g. Rossow, 1978; Lunine et al. 1989).
As temperature decreases, the photospheric density increases, more species condense so
that the particle density $\rho_p$ increases. Coalescence will become effective in growing large particles ($\tau_{coal}/\tau_{cond}\propto T/\rho_p^2$, Rossow 1978) and the afore-mentioned turbulent mixing could no longer prevent these large particles to fall out rapidly. Moreover the condensation line of some species (e.g. silicates) will drop below the photosphere (see e.g. Lodders 1999). Within this general scheme, the DUSTY models correspond to a situation with very efficient turbulent mixing, whereas the COND models correspond to a situation where mixing is inefficient compared with sedimentation. Whether this general mechanism, and the related decrease of opacity, can explain the color saturation - and thus smaller backwarming -
of the coolest L-dwarfs remains to be quantified correctly. Work in this direction is under progress.

As anticipated from Figure \ref{fig1}, spectral and photometric properties are
generally more
affected by dust formation than the inner atmospheric structure and thus the evolution, as shown in the next section. This reflects the weak dependence of $L$ and
$\te$ upon opacity for BDs, with $L\propto \kappa^{0.3}$ and
$\te\propto \sim \kappa^{0.1}$ (Burrows \& Liebert, 1993).

*** FIGURE 1 ***

\subsection{Evolutionary calculations}

Evolutionary calculations have been performed for $900\le \te \le 2800$ K,
which corresponds to masses between 0.1 $\msol$ and
 0.01 $\msol$ for ages $10^8\simle t\simle 10^{10}$ yr, following the method described in CB97. 
Figure 2 shows
the evolution of $\te$ as a function of time for various masses and dust treatments. This illustrates the uncertainty expected from
dust treatment on the cooling of BDs. For a given age, the differences
between the two extreme DUSTY and COND models
reach up to $\sim$10\% in $\te$ and $\sim$25\% in luminosity $L$. The hydrogen-burning minimum mass (HBMM) is only moderately affected by dust formation. The opacity due to the formation of different grains (e.g.  silicates, Al$_2$O$_3$, CaTiO$_3$) produces
a blanketing effect which lowers the effective temperature and luminosity for
a given mass at
 the bottom of the main sequence, as noted initially by Lunine et al
 (1989). Models
omitting grain opacity (NextGen and COND) yield 
$m_{\rm HBMM} \sim$ 0.072 $\msol$, $\te$=1700K, $\log L/\lsol = 5\times 10^{-5}$, whereas
DUSTY models yield a slightly lower limit, $m_{\rm HBMM} \sim$ 
0.07 $\msol$, $\te$=1550 K, $\log L/\lsol = 4\times 10^{-5}$, for solar composition. A lower luminosity requires a smaller
H-burning nuclear energy in the core
to reach thermal equilibrium, thus a lower central temperature and therefore a lower mass, which yields a lower
HBMM.

*** FIGURE 2 ***

The present calculations incorporate new conductive opacities by Potekhin et al. (1999) in the interior. These opacities, initially developed for neutron star envelopes and white dwarf cores, improve previous calculations (Hubbard and Lampe, 1969; Itoh et al. 1983 and Mitake et al., 1984; Brassard and Fontaine, 1994) and cover the range of temperatures and densities characteristic of BD interiors.
An interesting new property revealed by the present calculations is the
growth of a conductive core for the most massive BDs.
As these objects cool and contract, their interior becomes more and more degenerate
($\psi=kT/kT_F \approx 3.3 \times 10^{-6}\,T\,
(\mu_e/\rho)^{2/3}$,
where $T_F$ is the electron Fermi temperature and $\mu_e$ is the electron mean molecular weight).
Electron conductivity becomes more and more important and becomes eventually the main energy transport mechanism in the central regions, instead of convection. 
This occurs in the core of old ($t\ge$ 2 Gyr) BDs in the mass
range 0.03-0.07 $\msol$, with a maximum extension of the conductive core
$R_{cond}/R_\star\approx 0.6-0.7$ for the more massive ones, since the maximum central density occurs near the
HBMM (Burrows et al. 1997; Chabrier and Baraffe 2000). In this
mass range, the conductive core
appears at $t\approx$ 2-3 Gyr, for $\psi \simle 0.1$, and increases rapidly, reaching up to
50\%-70\% of the total mass at 10 Gyr for masses $m\approx 0.05-0.06$ $\msol$.  Figure \ref{fig3} shows the evolution of the conductive core $M_{cond}/M_\star$
as a function of time for a 0.06 $\msol$ BD.
We verified that these results still hold, although quantitatively different, when using previously published conductive opacities. These properties are essentially unaffected
by the dust treatment in the atmosphere, since evolution depends weakly on it. 

*** FIGURE 3 ***

The onset and growth of a conductive core does affect the evolution.
Once conduction sets in at the center, it increases dramatically the efficiency
of energy transport compared to convection ($v_F>>v_{conv}$, where $v_F=(2kT_F/m_e)^{1/2}$ is the electron Fermi velocity), decreasing the internal temperature gradient and yielding a cooler central temperature for a given energy flux
${\mathcal{F}}$ ($\nabla T^4\equiv T_c^4/R \propto {\mathcal{F}}. \kappa_{cond}$, where $\kappa_{cond}$ is the conductive opacity). This increases electron degeneracy
at the center ($\psi \propto kT$ decreases). In first approximation,
retaining only the zero-temperature electron contribution and the kinetic ionic contribution, the radius of a BD can be written $R\simeq R_0(1+\psi)$
where $R_0=2.8\times 10^9 ({m\over \msol})^{-1/3}\mu_e^{-5/3}$ cm is the zero-temperature (fully degenerate, $\psi=0$) radius (Stevenson, 1991).
Therefore the more degenerate conductive object 
contracts towards a slightly smaller radius, yielding a larger release of gravitational energy $\delta \Omega =\int_M P\delta(1/\rho) dm$ within a Kelvin-Helmoltz time.

This larger energy source increases the total binding energy of the star:

\begin{equation}
\delta B(t)=-[\delta \Omega(t)+\delta U(t)]=\int_M Gm\delta ({1\over r})dm\, -\int_M \delta \tilde udm
\end{equation}

\noindent where $\tilde u$ is the specific internal energy, yielding a larger luminosity $L$ for a given age,
i.e. slowing down the cooling time $\tau$ of the BD, since:

\begin{equation}
L(t)={dB(t)\over dt}\,\,\,\Rightarrow \tau(L)=\int_{t_0}^t {dB(t^\prime)\over L}
\end{equation}
\label{cool}

\noindent where we have neglected the nuclear source of energy arising from
some residual hydrogen-burning 
(see e.g. Figure 2 of Chabrier and Baraffe 2000).

As an example, for a 10 Gyr old 0.06 $\msol$ BD,
$\te$ is 15\% larger and $L$ is 30\% brighter when
conduction is taken into account. An effect difficult to verify observationally,
however, given the faintness of such old BDs.
Note that the onset of a conductive core occurs only for $t\simgr 1$ Gyr and is thus inconsequential for the application of the lithium-test, which concerns only younger objects (see e.g. CB97).

\section{Color - Magnitude diagrams}

In this section, we compare the present models with available observations of late-M, L- and methane-dwarfs in various CMDs.
The comparison is limited to objects with known distances. Although trigonometric parallaxes have been
determined for several 2MASS and DENIS L-dwarfs (Dahn et al. 2000),
the only methane dwarf with known distance is Gl229B.

\subsection{Optical CMD}

*** FIGURE 4 ***

Figure \ref{fig4} compares models and various observations in a $\mv$-$(V$-$I)$ CMD.
Three L-dwarfs of the Dahn et al. (2000) sample are known to be
very close binary systems (Mart\'\i n,  Brandner \& Basri 1999;
Koerner et al. 1999) with nearly equal mass components and
their magnitude is corrected in the CMDs to take into account this
property.
Models with different dust treatments are displayed
for ages of 0.1 Gyr (only for the DUSTY models; short-dash line) and 1 Gyr
(solid line). 
 For $t \, >$ 1 Gyr, DUSTY isochrones are hardly
distinguishable from one another in any of the following optical ($VRI$) and near-IR ($JHK$) CMDs,
although the same masses correspond obviously to different magnitudes,
and are not shown for sake of clarity.
The properties of the DUSTY models for different ages are displayed in Tables 1-5. Note that we do not give the H-magnitude. Indeed, the present models overestimate the flux in this band and work remains to be done to understand this shortcoming. We also give the lithium-abundance depletion for these models for applications of the
lithium-test. 
The COND models displayed in Figure 4 (long-dash line) cover the mass range between 0.08 $\msol$
($\te = 2350 K$) and 0.03 $\msol$ ($\te = 1050 K$) at 1 Gyr.
  
For $\mv$ brighter than 20 ($\te \simgr 2200$ K), the NextGen models (dash-dot) are
significantly bluer by 0.3-0.4 mag in $(V-I)$ 
than the DUSTY models. This is due to the recent improved TiO line
list (Schwenke, 1998) included in the DUSTY atmospheres (Allard et al. 2000a,b),
while the NextGen models used the previous line list by Jorgensen (1994). As
emphasized in BCAH98, the NextGen models predict 
too blue (up to $\sim 0.5$ mag) optical colors compared to observations for $10\simle \mv \simle 19$ (see Figure 6 of BCAH98).
As shown in Figure \ref{fig4}, the new TiO linelist improves significantly this situation (see also Allard et al. 2000b) and confirms
the fact that the previous discrepancy stemmed primarily from a lacking source of opacity shortward of 1 $\mu$m (cf. \S4.2 of BCAH98),
and not from a major caveat in the treatment of convection.
However, as shown
 in Figure \ref{fig4} and Figure \ref{fig5} below, the situation is not completely satisfactory yet, with a remaining
$\sim 0.2$-$0.3$ offset in colors at the bottom of the MS.

*** FIGURE 5 ***

The dust effect is clearly seen for $\mv \simgr 19$, $(V-I)  \simgr 4.8$, 
with a much more modest reddening as $\mv$ increases ($\te$ decreases) 
for the DUSTY and COND models than for
the dust-free models. This reflects metal and in particular TiO and VO depletion due to grain formation in the COND and DUSTY atmospheres. These species are two main sources of absorption in the optical in the temperature-range of interest. Note that, although the $(V-I)$ colors are similar, the DUSTY models are significantly fainter than the COND ones for a given mass, revealing the grain {\it opacity} which is not included in
COND models, and the depressed continuum in the optical due to the opacity-induced backwarming and H$_2$O dissociation (Allard et al. 2000a).
Interestingly enough, the same trend of change of slope in $(V-I)$ colors at $\mv \sim 19$ appears in the available observations, but at a $\sim 0.5$ mag bluer limit, i.e. $(V-I)\simle 4.5$,
than the theoretical predictions. This builds confidence in our basic treatment of grain formation but suggests that the gas depletion of the main absorbing species in the optical (TiO, VO, FeH) is underestimated in the present theory. Note in passing the limited interest of the V-band for BD search, with most
massive BDs older than 1 Gyr being at $\mv\simgr 24$.

The saturation in optical colors is even more drastic in $(R-I)$ (Figure \ref{fig5}), where
the DUSTY models yield almost constant $(R-I)$
below $\te \sim 2300$ K ($M_I\simgr$ 14) at $(R-I)\sim$2.4-2.5. As seen in the figure,
the distribution of various objects observed in these colors at the bottom of and below the main sequence, either in the field or in the Pleiades cluster,
not only shows the same saturation
effect as predicted by the models but is quantitatively in very good agreement with the predicted colors. As mentioned above, however, the $\sim 0.2-0.3$ mag offset above $\sim 2300$ K is still present.
The COND models predict bluer colors as $\te$ decreases, and 
in general more
flux in the $V,R,I$ bands at a given $\te$ than the DUSTY models.
This stems from (i) the absence of grain opacity in COND models and (ii) the backwarming effect in the DUSTY models (see Figure 1)
which destroys IR-absorbing species such as e.g. H$_2$, H$_2$O,
allowing more flux to escape at longer wavelengths.

Although the L3.5 dwarf
2MASS-WJ0036159+182 ($\mv$=21.15, $\mr$=18.15, $\mi$=15.92,
 $R$-$I$=2.23)(Reid et al. 1999) seems to be lying on the COND track, it is
probably misleading and reveals probable errors in the photometry. Indeed, the near-IR colors of this object are not 
reproduced by the COND models, but rather by the DUSTY ones (see Figure \ref{fig6}).

\subsection{Infra-red CMD}

Figure \ref{fig6} displays the same comparisons in near-IR colors $(J-K$). 
As already mentioned by Leggett et al. (1998), the DUSTY models give
the best fit to IR photometry of L- and very late M-dwarfs. They explain
the very red colors characteristic of these objects in terms of the backwarming
of the atmosphere due to grain absorption, yielding a
reduction of the band strengths of the main IR absorbers, like e.g. H$_2$O,
as proposed initially by Tsuji et al. (1996). 
Such red colors are 
impossible to reproduce with dust-free opacity models (NextGen and COND).
Note that for $\mk$ brighter than $\sim$ 11, the DUSTY models are slightly bluer
in $(J-K)$, by $\sim$0.1-0.2 mag, than the NextGen models. This arises from
the use of different H$_2$O
line lists: Miller et al. (1994) in NextGen and the recent AMES list (Partridge and Schwenke, 1997)
in DUSTY models. Although less complete for the higher energy transitions
than the AMES line list, the Miller et al. (1994) list yields better agreement
between models and IR photometric observations for $\te \simgr 2200$-$2300$ K (where dust does not affect the colors), since it relies on better potential surfaces for high temperatures (see Allard et al. 2000b for details). Below $\te \sim 2000$ K, these transitions are no longer important and the more complete AMES water linelist is to be used. Below this temperature, however, dust opacity
is the main ingredient which shapes the IR colors, and the choice of the
water line list is less crucial for photometric analysis. This illustrates again the remaining shortcomings in the present theory arising from still partially inaccurate molecular opacities, and stresses the need for future improvements in these inputs.
 
*** FIGURE 6 ***

When dust opacity is neglected (COND and NextGen), 
near-IR colors become bluer due to strong H$_2$, H$_2$O, CO and eventually CH$_4$ absorption in the $H$ and $K$-bands.
This yields very blue near-IR colors, with $J-K$$\sim $$J-H\simle 0.5$, as illustrated by
the good agreement of Gl229B IR  colors with the COND track (see also Burrows et
al. 1997).
The effect is more pronounced for the COND models because of the lack of TiO absorption, which extends into the J-band at these very low effective temperatures.
However, the COND models
overestimate the flux at and shortward of $\sim 1\, \mu m$ ($V,R,I$ bands) for Gl229B (Golimowski et al. 1998).
The spectrum of GL229B indeed does not show strong indication for dust
in its IR spectrum from 1 to 5 $\mu  m$ (Allard et al. 1996; Marley et al. 1996; Tsuji et al. 1996; Oppenheimer et al. 1998),
but reveals discrepancies in the optical with dust-free models,
as mentioned above.
Several solutions have been proposed for this missing source
of opacity in the optical, such as a specific haze opacity due to
the presence of the nearby star (Griffith et al. 1998) or warm dust layers (Tsuji et al. 1999). These suggestions, however,
are based on ad-hoc hypothesis required to fit Gl229B spectrum and lack robust
physical grounds. The methane-dwarf SDSS 1624 (Strauss et al. 1999), for example, does not
have a hot companion, yet shows 
no flux blueward of 0.8 $\mu$m, indicating
a strong absorption in the optical, as does Gl229B. This suggests that the source of absorption is a common atmospheric
BD property, precluding the haze opacity scenario.
Strong alkali resonance lines (K I, Na I) have been proposed recently as an alternative, or complementary solution (Tsuji et al. 1999; Burrows et al. 1999).
Clearly this question is not settled yet and requires futher observational tests and consistent calculations.

For $(J-K) \sim$ 1.0-1.9, good agreement is found
between the data and the DUSTY models (see Figure \ref{fig6}), except for the L0-dwarf 2MASS-WJ0746425+200032
(Reid et al. 1999) ($\mk$=9.70, $J$-$K$=1.25),
which lies significantly above both the other data and the models, making it a probable unresolved binary, as suggested by these authors.

An interesting feature of this diagram is a hint that the redder L-dwarfs
of Dahn et al. (2000), and the slope of the observed L-dwarf sequence itself, seem to deviate from the DUSTY tracks for $\mk\simgr 11.5$ and $(J-K) \simgr
1.5$, the reddest observed objects having a color $(J-K)\sim 1.9$, i.e. $\te \sim$ 1800 K.
If the parallax and the photometry
of these objects are confirmed,
this may very well illustrate gravitational settling of some grains below the photosphere
for these objects, as mentioned in \S2.1. This will yield less backwarming and thus progressively cooler atmospheres (see Figure 1), increasing H$_2$O, CO and H$2$ absorption in the IR.
Once CH$_4$ forms near the photosphere, the integrated flux over the K-band is strongly reduced, although the flux still peaks around 1 $\mu$m, leading to Gl229B-like colors. Whether this transition occurs sharply or smoothly remains to be elucidated. Note that the CO-CH$_4$ transition occurs gradually with some of the two elements being present in the stability field of the other so that cool stars do contain some limited amount of methane and CO has been detected in Gl229B (Noll et al. 1997; Oppenheimer et al. 1998). The recently discovered methane-dwarfs (Burgasser et al. 1999), unfortunately, do not bring information about this transition because of their color selection criterium, $J-K<0.4$. In case of an abrupt change in color between "L" and "methane"-dwarfs,
there should be indeed a "no-BD land" in the region $0.5\simle J-K\simle 1.0$
with a sharp transition for $\te\sim T_{\rm gas}=  1300-1400$ K, the temperature of formation of methane in the $P\approx 3-10$ bar range. This corresponds to masses $m\approx 0.015-0.02\,\msol$ for t=0.1 Gyr, $m\approx 0.045-0.05\,\msol$  for t=1 Gyr and $m\approx 0.065\,\msol$ for t=5 Gyr. In any events, this concerns objects cooler than $\te\approx $ 1700-1800 K, the temperature derived with the DUSTY models along the 1 Gyr isochrone for LHS102B
(Goldman et al. 1999; see also Figure 6), the reddest L-dwarf
with known parallax, and for 2MASS-W1632+19,
the faintest L-dwarf in the Dahn et al. (2000) sample with $\mk$ $\sim$ 12.75, $(J-K$) $\sim$ 1.9. These temperatures agree within 100 K with the ones
derived spectroscopically
 by Basri et al. (1999) based on
alkali resonance absorption lines in the optical spectrum. Objects cooler
than 2MASS-W1632+19 thus correspond to BDs with masses 
$m\simle 0.03$ $\msol$ for ages $t>0.1$ Gyr, $m\simle 0.06$ $\msol$ for $t>1$ Gyr and to essentially {\it all} BDs for $t\simgr 2.5$ Gyr (see Figure \ref{fig2}).

The same general behavior as in ($J-K$) is found in ($J-H$), with
the same color saturation observed for the reddest L dwarfs for $\mj \simgr 13$
and $(J-H)\simle 1.2$. Although not shown
in the present paper, a COND isochrone at $t\sim$(a few)$\times$10$^8$ yr fits 
well GL229B both in $(J-H$), in contrast to
Burrows et al. (1997) models, and $(J-K$) colors. Their models could not reproduce the $L$'-band of GL229B.
Since then, Leggett et al. (1999) revised the flux of GL229B, with a significant
modification of the $L$' band, which is now reproduced by the Burrows et al. (1997)
models. The reason of their poor fit in the H-band, however, compared with the present models, remains unclear. Indeed the explanation proposed in Burrows et al. (1997) was based on previous, less accurate observations of Gl229B, as mentioned above, and does not hold anymore.
 
Finally, Figure 7 displays $L'$ versus $(K-L'$). The $K$ band is affected by H$_2$, H$_2$O and
CH$_4$ absorption, whereas the $L'$ is strongly affected by methane  
(Allard et al. 1996; Marley et al. 1996; Tsuji et al. 1996). The trends displayed in the previous IR diagrams are replicated, with a severe reddening due to grain backwarming effects in the DUSTY models, but also a substantial reddening for the COND and grainless model, due to the strong CH$_4$ absorption
in the K-band.

The magnitudes of brown dwarfs and very-low-mass stars with dusty atmospheres for several isochrones are displayed in Tables 1-5. 
We stress, however, that
the flux for these objects peaks around 1 $\mu m$, so that J,Z and H remain the most favorable bands for detection.
 
*** FIGURE 7 ***

\section{Discussion and conclusion}

The comparisons in various CMDs displayed in the previous section illustrate the good general agreement between the present theory of evolution of very-low-mass stars and brown dwarfs with dusty atmospheres and observations.
The models which include grain formation
{\it and} opacity in the atmosphere explain successfully the IR colors of late M- and L-dwarfs, as well as their blue loop in the optical, although improvement
is still needed for the latter case to reach accurate quantitative agreement with
observations.
As shown by Kirkpatrick et al. (1999a) for GD165B and Ruiz et al. (1997) for Kelu-1, the DUSTY models improve also significantly the agreement with observed IR spectra. The present models
provide a consistent description of the physics going on in the interior and the atmosphere of objects at the bottom of and below
the main sequence, from M-dwarfs to L-dwarfs. They can be used
with reasonable confidence in the IR
bands ($JKL'M$) to calibrate the main properties, mass, age, $\te$, $L$ of identified late-M and L-dwarfs. 
Shortcomings still appear in the optical, whatever the treatment of dust, in spite of noticeable improvement (see Figure 4). Baraffe 
et al. (1998) already pointed out a possible missing source of opacity in the optical
to explain the observed $(V-I)$ colors of M-dwarfs between $\te \sim$ 3600 K and $\sim$ 2300 K.
Although the new TiO linelist from AMES (Schwenke, 1998) improves appreciably the situation
(see Figure \ref{fig4}), a $\sim 0.2-0.3$ mag discrepancy still remains with observations (see Figure \ref{fig5}).
Below $\sim$ 2300 K, comparison of the DUSTY models with observations
in $(V-I)$ (see Figure \ref{fig4}) strongly suggests either an underestimate of
optical opacities or an underestimate of the depletion of
molecular absorbers during grain condensation. Problems appear as well in spectroscopic analysis of  atomic line profiles 
where the DUSTY models cannot match
the optical observations (Tinney et al. 1998; Basri et al. 1999).
Another remaining caveat of the present theory arises from shortcomings
in the absorption coefficients of water, 
yielding a $\sim 0.2$ mag offset in $(J-K)$ for $\te \simgr 2300$ K
when using the most recent H$_2$O line list
(see e.g. Figure 6).
Finally,
the optical spectrum of GL229B remains to be described accurately. Present
COND synthetic spectra
overestimate the optical flux, pointing out a missing
source of opacity. Work in this direction is under progress.

Although it is important to solve these problems and eliminate these shortcomings, it is obvious from the various CMDs that
substellar objects must be searched in the near-infrared. Brown dwarfs around 1500 K radiate more than 90\% of their
energy at wavelengths longward of $1 \, \mu$m. The flux peaks at 1.1 and 1.3 $\mu$m, and $J,Z,H$ are the favored broadband filters for detection.
It is thus reinsuring, given the complexity of the underlying physics, that the present models of dusty substellar objects
describe the observations in near-IR bands with reasonable success.

If dust formation and absorption successfully explain the transition between M- and
L-dwarfs, a theoretical and observational gap still exists between L- and methane-dwarfs.
Models including dust opacity reach rapidly very red IR colors ($J-K$) $>$ 2,
whereas observations of the reddest or faintest L dwarfs show a saturation
of their ($J-K$) and ($J-H$) colors at 1.9 and 1.2 respectively, which corresponds to $\te$
$\sim$ 1700-1800 K (see Figure \ref{fig6}). This very likely illustrates ongoing dynamical processes near the photosphere.
As outlined in \S2.1, the proximity of the top of the convection zone from the photosphere
may induce turbulent mixing and prevent the grains to diffuse downward. As convection retreats to deeper layers for cooler objects, gravitational settling of condensed species may lead to cooler atmospheres, because of the decreasing opacity, with a transition to very blue near infrared colors once methane forms near the photosphere. 
These
dynamical processes of grain formation and diffusion are not included in the
theory yet and represent indeed the next major challenge. Theory
definitely needs some guidance from observations through this puzzle, by
getting spectra, parallaxes and photometry of the coolest L-dwarfs and by either filling or confirming the presently expected brown dwarf dearth between the coolest L-dwarfs and 
the hottest methane dwarfs around $\te \approx 1300-1400$ K.

\medskip
{\it Note: }
Various isochrones for the DUSTY models from 10$^7$ yr to 10$^{10}$
yr, covering a range
of mass from 0.01 $\msol$ to 0.1 $\msol$ and $\te$ from 900K to 3000K
are available by anonymous ftp
(same format as Table 1):
\par
\hskip 1cm ftp ftp.ens-lyon.fr \par
\hskip 1cm username: anonymous \par
\hskip 1cm ftp $>$ cd /pub/users/CRAL/ibaraffe \par
\hskip 1cm ftp $>$ get DUSTY00\_models \par
\hskip 1cm ftp $>$ quit
\bigskip

Note that the DUSTY models are given in the Tables down to 
$\te$=900K, but this limit is clearly unrealistic for dusty atmospheres, as seen in the various figures of the paper.

Special requests for models in any
particular filters can be addressed to I. Baraffe.

\begin{acknowledgements} We are very grateful to C. Dahn and H. Harris for providing data prior to
publication and to A. Potekhin for providing the conductive opacities.
We are also thankful to E. Mart\'\i n and to the referee, D. Saumon, for constructive suggestions and comments.
 Part of this work was performed at the AAO (Sidney) 
  under the auspice of the franco-australian cooperation program
  in Astronomy (CNRS/ARC).
 G.C and I.B are very indebted to B. Boyle, C. Tinney and the staff of the AAO for their generous hospitality during this visit.

This work was supported in part by a CNRS-NSF "dark matter" grant, NSF grant
AST-9720704, NASA ATP grant NAG 5-3018 and LTSA grant NAG 5-3619 to the
University of Georgia, and NASA LTSA grant NAG5-3435 to Wichita State
University. The computations were done at the P\^ole Scientifique de
Mod\'elisation Num\'erique at ENS-Lyon, on the T3E of Centre
d'Etudes Nucl\'eaires de Grenoble, on the IBM SP2 of CNUSC, the SGI Origin 2000 of the
UGA UCNS, on the IBM SP2 of the San Diego Supercomputer Center (SDSC) with
support from the National Science Foundation, and on the Cray T3E of the NERSC
with support from the DoE. We thank all these institutions for a generous
allocation of computer time.
\end{acknowledgements}

\clearpage\eject

\petit{

\begin{table*}
\caption{DUSTY model isochrone for 0.1 Gyr. 
$T_{eff}$ is in K, the gravity $g$ in cgs.
The lithium abundance
is normalized to the initial abundance, which is $X_{Li_0} = 10^{-9}$ by mass fraction in
the present calculations.
 The VRI magnitudes are in the 
Johnson-Cousins system (Bessell 1990), JHK in the CIT system (Leggett 1992), L$^\prime$ in the Johnson-Glass system and M in the Johnson system. 
}
\begin{tabular}{lcccccccccccc}
\hline\noalign{\smallskip}
$m/\msol$  &$T_{eff}$ & $\log L/\lsol$ & $\log \,g$ & $R/R_\odot$ & Li/Li$_0$ &$M_V$ &$M_R$ &$M_I$ &$M_J$ &
$M_K$ & $M_{L'}$ & $M_M$\\
\noalign{\smallskip}
\hline\noalign{\smallskip}
 0.010&  916. & -4.98& 4.21& 0.130& 1.0000 & 42.64& 35.06& 31.45& 22.06&  15.20& 11.71& 11.30\\
 0.012& 1333. & -4.25& 4.22& 0.141& 1.0000 & 29.33& 24.92& 22.13& 16.50&  12.14& 10.10& 10.12\\
 0.015& 1290. & -4.37& 4.37& 0.132& 1.0000 & 30.63& 25.96& 23.08& 17.14&  12.52& 10.32& 10.36\\
 0.020& 1441. & -4.21& 4.53& 0.127& 1.0000 & 27.30& 23.35& 20.73& 15.68&  11.86& 10.14& 10.20\\
 0.030& 1878. & -3.75& 4.71& 0.127& 1.0000 & 21.21& 18.37& 15.96& 12.22&  10.76&  9.74&  9.89\\
 0.040& 2234. & -3.41& 4.80& 0.132& 1.0000 & 18.72& 16.41& 13.95& 11.04&  10.16&  9.37&  9.62\\
 0.050& 2483. & -3.18& 4.84& 0.140& 0.9990 & 17.07& 15.14& 12.86& 10.47&   9.70&  9.04&  9.28\\
 0.055& 2576. & -3.09& 4.86& 0.145& 0.9920 & 16.44& 14.65& 12.48& 10.27&    9.51&  8.90&  9.12\\
 0.060& 2652. & -3.01& 4.87& 0.149& 0.9630 & 15.95& 14.26& 12.18& 10.11&    9.36&  8.77&  8.98\\
 0.070$^\star$& 2768. & -2.88& 4.88& 0.159& 0.6680 & 15.19& 13.64& 11.70&  9.83&   9.09&  8.55&  8.72\\
 0.072& 2787. & -2.86& 4.88& 0.161& 0.5380 & 15.07& 13.54& 11.63&  9.78&   9.04&  8.51&  8.68\\
 0.075& 2814. & -2.83& 4.88& 0.164& 0.3290 & 14.89& 13.40& 11.52&  9.71&   8.97&  8.45&  8.62\\
 0.080& 2854. & -2.78& 4.89& 0.169& 0.0664 & 14.65& 13.19& 11.36&  9.61&    8.87&  8.36&  8.52\\
 0.090& 2920. & -2.69& 4.89& 0.178& 0.0000 & 14.23& 12.84& 11.09&  9.41&    8.68&  8.20&  8.34\\
 0.100& 2971. & -2.63& 4.90& 0.185& 0.0000 & 13.92& 12.58& 10.89&  9.27&    8.54&  8.07&  8.21\\
\hline
\end{tabular}
\begin{list}{}{}
\item[$^\star$] HBMM
\end{list}
\end{table*}

\begin{table*}
\caption{Same as Table 1 for 0.5 Gyr
}
\begin{tabular}{lcccccccccccc}
\hline\noalign{\smallskip}
$m/\msol$  &$T_{eff}$ & $\log L/\lsol$ & $\log \,g$ & $R/R_\odot$ & Li/Li$_0$ &$M_V$ &$M_R$ &$M_I$ &$M_J$ &
 $M_K$ & $M_{L'}$ & $M_M$\\
\noalign{\smallskip}
\hline\noalign{\smallskip}
 0.020&  901. & -5.14& 4.65& 0.111& 1.0000 & 43.80& 35.89& 32.23& 22.56& 15.59&
12.11& 11.67\\
 0.030& 1170. & -4.73& 4.86& 0.106& 1.0000 & 33.93& 28.57& 25.49& 18.78& 13.65&
11.09& 11.10\\
 0.040& 1461. & -4.36& 5.01& 0.104& 1.0000 & 27.41& 23.50& 20.90& 15.94& 12.20&
10.51& 10.62\\
 0.050& 1751. & -4.06& 5.11& 0.103& 0.9540 & 22.45& 19.47& 17.10& 13.18& 11.35&
10.23& 10.36\\
 0.055& 1902. & -3.91& 5.15& 0.103& 0.5610 & 21.42& 18.64& 16.25& 12.60& 11.20&
10.16& 10.32\\
 0.060& 2048. & -3.78& 5.19& 0.104& 0.0069 & 20.56& 17.91& 15.46& 12.12& 11.00&
10.04& 10.25\\
 0.070$^\star$& 2295. & -3.56& 5.23& 0.106& 0.0000 & 18.95& 16.62& 14.18& 11.42& 10.58&
 9.78& 10.06\\
 0.072& 2332. & -3.52& 5.24& 0.107& 0.0000 & 18.71& 16.44& 14.01& 11.34& 10.51&
 9.74& 10.02\\
 0.075& 2400. & -3.46& 5.24& 0.108& 0.0000 & 18.26& 16.09& 13.71& 11.18& 10.39&
 9.65&  9.93\\
 0.080& 2505. & -3.37& 5.25& 0.111& 0.0000 & 17.56& 15.56& 13.27& 10.95& 10.18&
 9.51&  9.79\\
 0.090& 2678. & -3.19& 5.24& 0.118& 0.0000 & 16.40& 14.67& 12.58& 10.57&  9.83&
 9.24&  9.48\\
 0.100& 2809. & -3.05& 5.23& 0.128& 0.0000 & 15.52& 13.97& 12.06& 10.25&  9.52&
 8.99&  9.19\\

\hline
\end{tabular}
\begin{list}{}{}
\item[$^\star$] HBMM
\end{list}
\end{table*}

\begin{table*}
\caption{Same as Table 1 for 1 Gyr 
}
\begin{tabular}{lcccccccccccc}
\hline\noalign{\smallskip}
$m/\msol$  &$T_{eff}$ & $\log L/\lsol$ & $\log \,g$ & $R/R_\odot$ & Li/Li$_0$ &$M_V$ &$M_R$ &$M_I$ &$M_J$ &
 $M_K$ & $M_{L'}$ & $M_M$\\
\noalign{\smallskip}
\hline\noalign{\smallskip}
 0.030&  958. & -5.11& 4.90& 0.102& 1.0000 & 40.76& 33.66& 30.17& 21.50&  15.22& 11.98& 11.68\\
 0.040& 1189. & -4.77& 5.06& 0.098& 1.0000 & 33.53& 28.29& 25.25& 18.69&  13.69& 11.20& 11.24\\
 0.050& 1424. & -4.48& 5.18& 0.096& 0.9320 & 28.26& 24.20& 21.54& 16.42&  12.53& 10.74& 10.86\\
 0.055& 1550. & -4.34& 5.22& 0.095& 0.3790 & 26.08& 22.48& 20.00& 15.39&  12.08& 10.59& 10.71\\
 0.060& 1675. & -4.21& 5.27& 0.094& 0.0000 & 23.95& 20.72& 18.33& 14.14&  11.71& 10.48& 10.59\\
 0.070$^\star$& 2012. & -3.88& 5.32& 0.096& 0.0000 & 21.10& 18.37& 15.95& 12.47&  11.24& 10.24& 10.43\\
 0.072& 2089. & -3.80& 5.32& 0.097& 0.0000 & 20.51& 17.87& 15.41& 12.17&  11.10& 10.15& 10.37\\
 0.075& 2185. & -3.71& 5.33& 0.098& 0.0000 & 19.82& 17.32& 14.85& 11.85&  10.92& 10.03& 10.29\\
 0.080& 2358. & -3.54& 5.32& 0.103& 0.0000 & 18.66& 16.41& 14.00& 11.38&  10.57&  9.80& 10.08\\
 0.090& 2625. & -3.27& 5.28& 0.113& 0.0000 & 16.81& 14.99& 12.84& 10.74&   9.99&  9.38&  9.64\\
 0.100& 2790. & -3.08& 5.25& 0.125& 0.0000 & 15.68& 14.10& 12.16& 10.33&   9.59&  9.05&  9.26\\
\hline
\end{tabular}
\begin{list}{}{}
\item[$^\star$] HBMM
\end{list}
\end{table*}

\begin{table*}
\caption{Same as Table 1 for 5 Gyr 
}
\begin{tabular}{lcccccccccccc}
\hline\noalign{\smallskip}
$m/\msol$  &$T_{eff}$ & $\log L/\lsol$ & $\log \,g$ & $R/R_\odot$& Li/Li$_0$ &$M_V$ &$M_R$ &$M_I$ &$M_J$ &
 $M_K$ & $M_{L'}$ & $M_M$\\
\noalign{\smallskip}
\hline\noalign{\smallskip}
 0.050&  907. & -5.35& 5.26& 0.086& 0.9210 & 43.03& 35.38& 31.77& 22.53&  15.95& 12.57& 12.11\\
 0.055&  990. & -5.21& 5.32& 0.085& 0.2840 & 39.43& 32.76& 29.36& 21.21&  15.26& 12.23& 11.95\\
 0.060& 1097. & -5.04& 5.37& 0.084& 0.0000 & 35.60& 29.92& 26.76& 19.74&  14.50& 11.81& 11.74\\
 0.070$^\star$& 1593. & -4.36& 5.40& 0.087& 0.0000 & 25.54& 22.08& 19.67& 15.23&  12.13& 10.73& 10.84\\
 0.072& 1754. & -4.18& 5.40& 0.089& 0.0000 & 22.96& 19.93& 17.58& 13.63& 
11.68& 10.54& 10.66\\
 0.075& 1998. & -3.92& 5.38& 0.092& 0.0000 & 21.31& 18.55& 16.14& 12.62&  11.35& 10.32& 10.50\\
 0.080& 2295. & -3.61& 5.34& 0.100& 0.0000 & 19.12& 16.77& 14.32& 11.57&  10.73&  9.91& 10.20\\
 0.090& 2621. & -3.28& 5.29& 0.113& 0.0000 & 16.85& 15.02& 12.86& 10.75& 10.01&  9.39&  9.65\\
 0.100& 2791. & -3.08& 5.25& 0.124& 0.0000 & 15.68& 14.10& 12.16& 10.33&    9.59&  9.05&  9.26\\
\hline
\end{tabular}
\begin{list}{}{}
\item[$^\star$] HBMM
\end{list}
\end{table*}

\begin{table*}
\caption{Same as Table 1 for 10 Gyr 
}
\begin{tabular}{lcccccccccccc}
\hline\noalign{\smallskip}
$m/\msol$  &$T_{eff}$ & $\log L/\lsol$ & $\log \,g$ & $R/R_\odot$& Li/Li$_0$ &$M_V$ &$M_R$ &$M_I$ &$M_J$ &
 $M_K$ & $M_{L'}$ & $M_M$\\
\noalign{\smallskip}
\hline\noalign{\smallskip}
 0.060&  971. & -5.29& 5.40& 0.081& 0.0000 & 40.34& 33.44& 30.00& 21.61&  15.52& 12.43& 12.09\\
 0.070$^\star$& 1552. & -4.42& 5.41& 0.086& 0.0000 & 26.27& 22.67& 20.19& 15.60&  12.28& 10.79& 10.91\\
 0.072& 1744. & -4.19& 5.40& 0.089& 0.0000 & 23.10& 20.05& 17.69& 13.71& 
11.70& 10.55& 10.67\\
 0.075& 1998. & -3.92& 5.38& 0.092& 0.0000 & 21.31& 18.55& 16.14& 12.62&  11.35& 10.32& 10.50\\
 0.080& 2297. & -3.61& 5.34& 0.100& 0.0000 & 19.11& 16.76& 14.31& 11.56&  10.72&  9.91& 10.20\\
 0.090& 2622. & -3.27& 5.29& 0.113& 0.0000 & 16.84& 15.01& 12.85& 10.75&  10.01&  9.39&  9.65\\
 0.100& 2793. & -3.08& 5.25& 0.124& 0.0000 & 15.66& 14.09& 12.15& 10.33&   9.59&  9.05&  9.26\\

\hline
\end{tabular}
\begin{list}{}{}
\item[$^\star$] HBMM
\end{list}
\end{table*}

}

\vfill

\clearpage\eject

\begin{figure}

\centerline {\bf FIGURE CAPTIONS}
\vskip1cm

\caption[]{Atmosphere profiles for $\te$=1800 K 
and $\log \, g$=5. Different treatments of dust are considered. Solid line (full circle):
 DUSTY; long-dashed line (full square): COND; dash-dotted
line (open circle): NextGen (see text). The photosphere ($\tau_{\rm Rosseland} \sim 1$)
is located at $T \sim 2100$ K for all models, whereas the end of the curves correspond to $\tau=100$, the boundary condition with the interior profile. The symbols on each curve indicate
the onset of convection}
\label{fig1}
\end{figure}

\begin{figure}
\caption[]{Evolution of $\te$ as a function of time (in yr) for different
masses and atmosphere models. Solid lines: DUSTY; long-dashed lines: COND; 
dash-dotted lines: NextGen. Masses are indicated in $\msol$}
\label{fig2}
\end{figure}

\begin{figure}
\caption[]{Evolution of the conductive core $M_{\rm cond}/M_{\rm tot}$ (shaded area) as a function of
time for a 0.06 $\msol$ brown dwarf.}
\label{fig3}
\end{figure}

\begin{figure}
\caption[]{CMD for the optical colors $(V-I)$ at the bottom of the Main Sequence
and below. The data is from Monet et al. (1992) (filled circles),
Dahn et al. (1995) (crosses),
Dahn et al. (2000) (filled squares) and Reid et al. (1999) (filled triangles).
Kelu-1 (Ruiz et al. 1997) is indicated by a full diamond.
Theoretical models correspond to various isochrones: 
short-dashed and solid lines: DUSTY models for 0.1 and 1 Gyr respectively;
long-dashed line: COND models for 1 Gyr; 
dash-dotted line: NextGen models for 1 Gyr. Masses (in $\msol$) and $\te$
are indicated on the 1 Gyr DUSTY isochrone by open circles.
The COND models go from 0.08 $\msol$ (same as DUSTY) to 0.03 $\msol$.
See Tables 1-5 for the location of the HBMM at different ages}
\label{fig4}
\end{figure}

\begin{figure}
\caption[]{CMD for the optical color $(R-I)$.
 The filled circles are the Pleiades data from Bouvier et al. (1998).
The other symbols and the models are the same as in Figure \ref{fig4}. 
See Tables 1-5 for the location of the HBMM at different ages}
\label{fig5}
\end{figure}

\begin{figure}
\caption[]{CMD for the near-IR color $(J-K)$.
 The dots are the data from Leggett (1992).
The other symbols are the same as in Figure \ref{fig4}.
Some identified BDs are indicated by diamonds: Kelu-1 (Ruiz et al. 1997); 
 GD165B (Kirkpatrick et al. 1999a); LHS102B (Goldman et al. 1999).
Theoretical models are the same as in Fig \ref{fig4}.
Crosses on the COND track indicate masses in $\msol$ and $\te$ in K.
See Tables 1-5 for the location of the HBMM at different ages}
\label{fig6}
\end{figure}

\begin{figure}
\caption[]{CMD for the IR color $(K-L')$.
The dots are data from Leggett (1992). 
Theoretical models are the same as in Figure \ref{fig6}, with the same
masses indicated on the COND isochrone (long-dash line, $\times$).
The dotted curve corresponds to the DUSTY isochrone for 10 Gyr.
 Masses in $\msol$
are indicated on the DUSTY isochrone for 1 Gyr (solid line, O); $\te$ for
$m \ge 0.08 \msol$ are the same as in Figure \ref{fig4}.
See Tables 1-5 for the location of the HBMM at different ages}
\label{fig7}
\end{figure}

\vfill\eject
\begin{figure}
\begin{center}
\epsfxsize=180mm
\epsfysize=180mm
\epsfbox{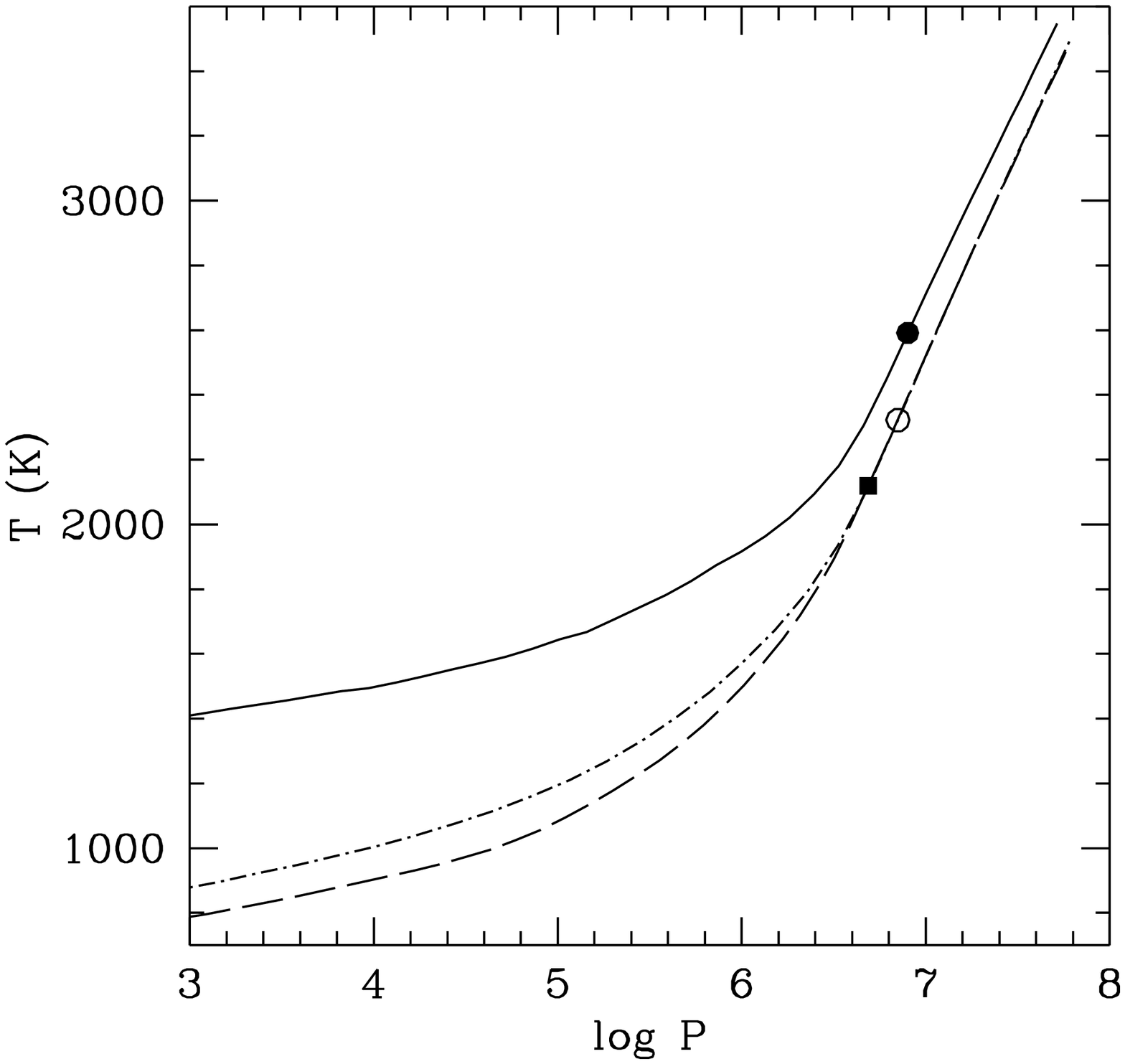}
\end{center}
\end{figure}

\vfill\eject
\begin{figure}
\begin{center}
\epsfxsize=180mm
\epsfysize=180mm
\epsfbox{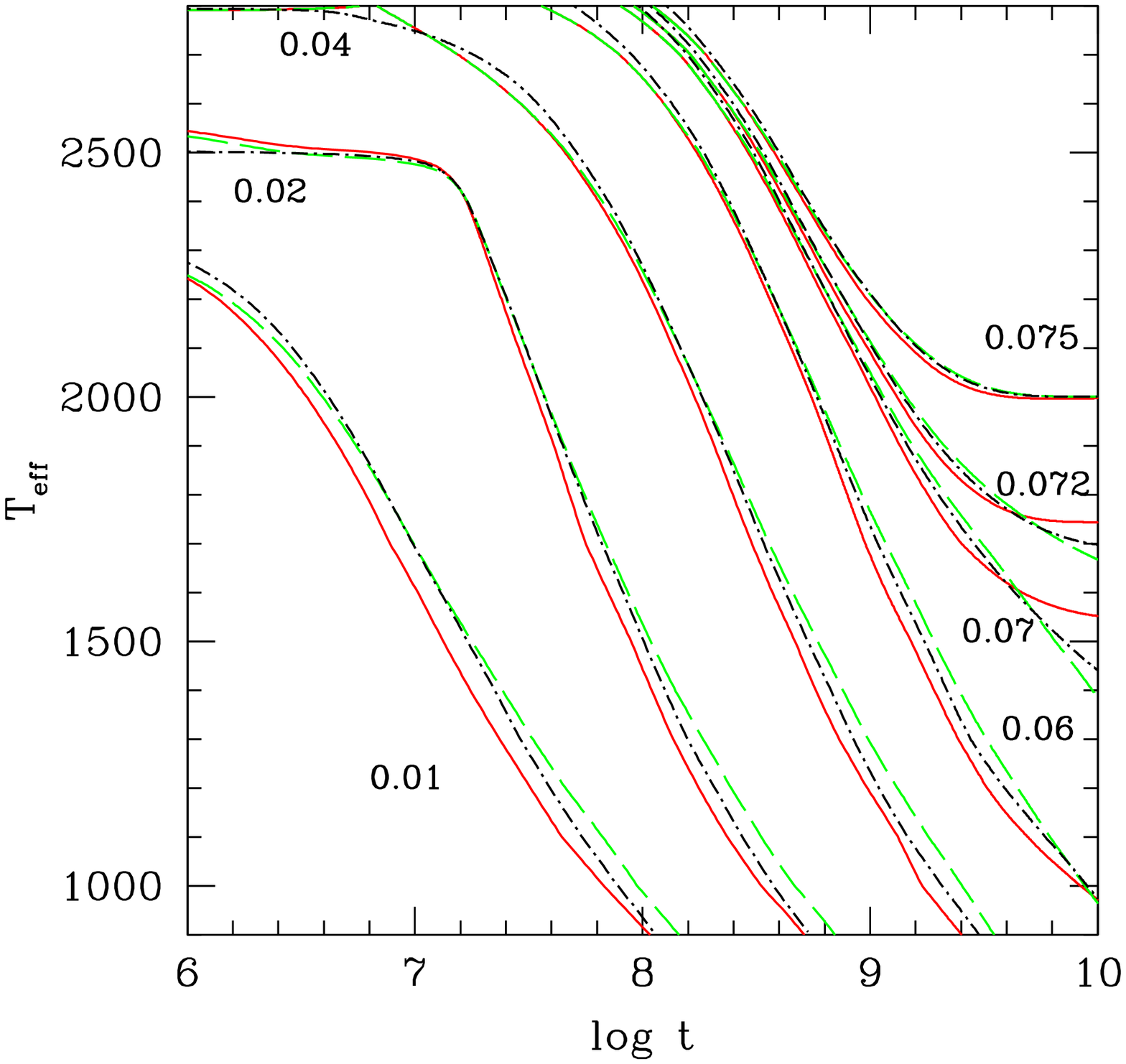}
\end{center}
\end{figure}

\vfill\eject
\begin{figure}
\begin{center}
\epsfxsize=180mm
\epsfysize=180mm
\epsfbox{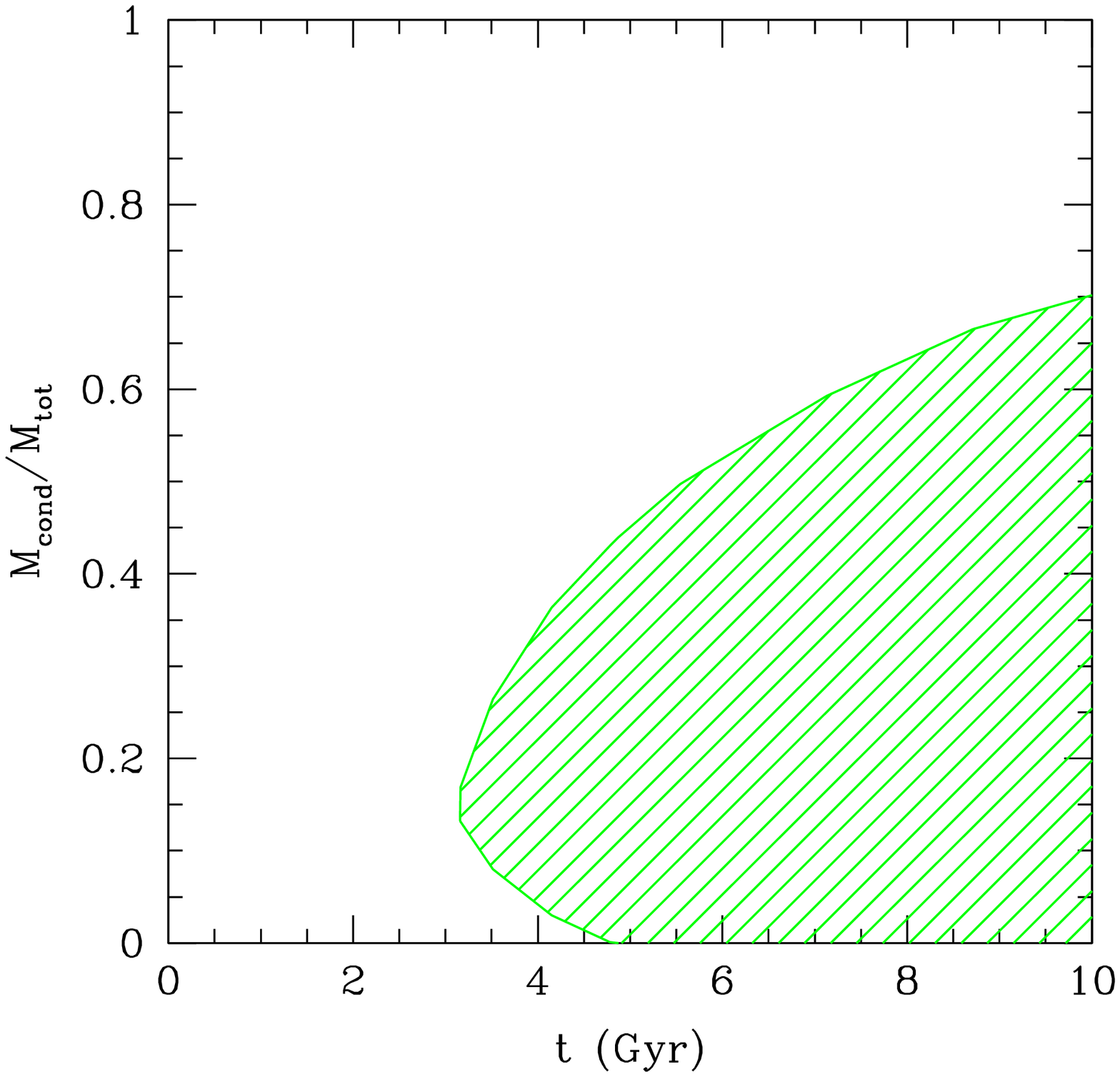}
\end{center}
\end{figure}

\vfill\eject
\begin{figure}
\begin{center}
\epsfxsize=180mm
\epsfysize=180mm
\epsfbox{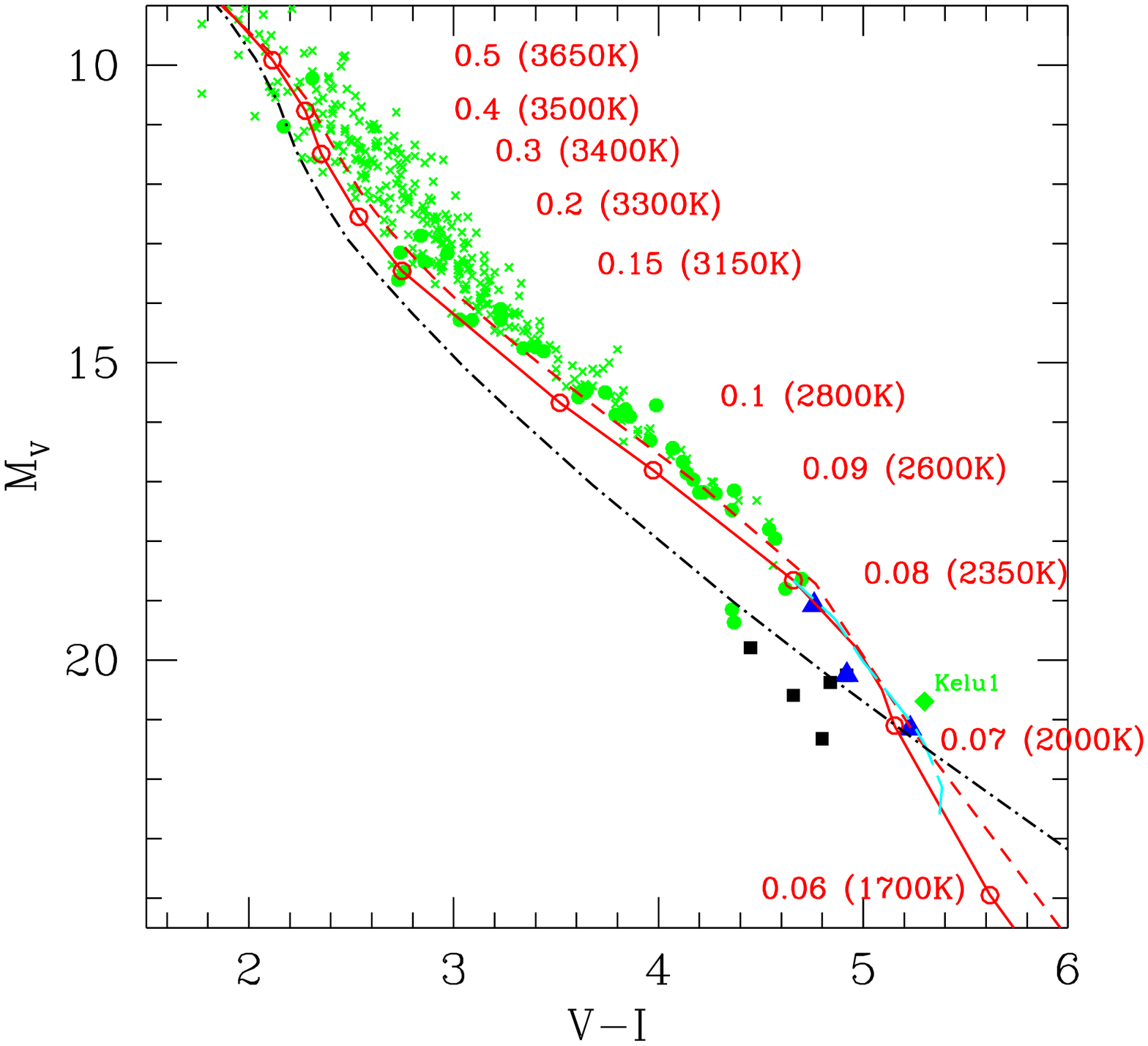}
\end{center}
\end{figure}

\vfill\eject
\begin{figure}
\begin{center}
\epsfxsize=180mm
\epsfysize=180mm
\epsfbox{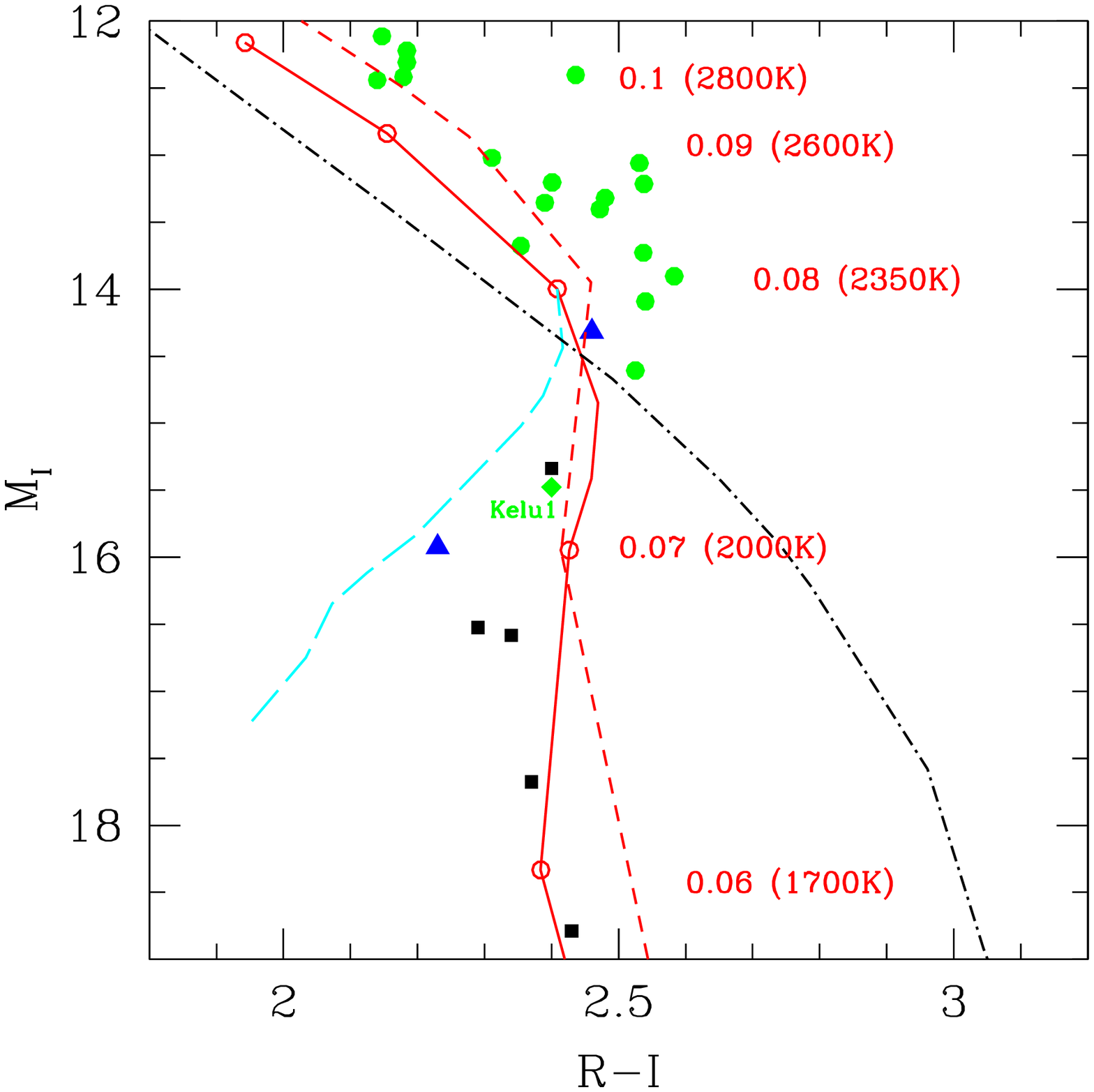}
\end{center}
\end{figure}

\vfill\eject
\begin{figure}
\begin{center}
\epsfxsize=180mm
\epsfysize=180mm
\epsfbox{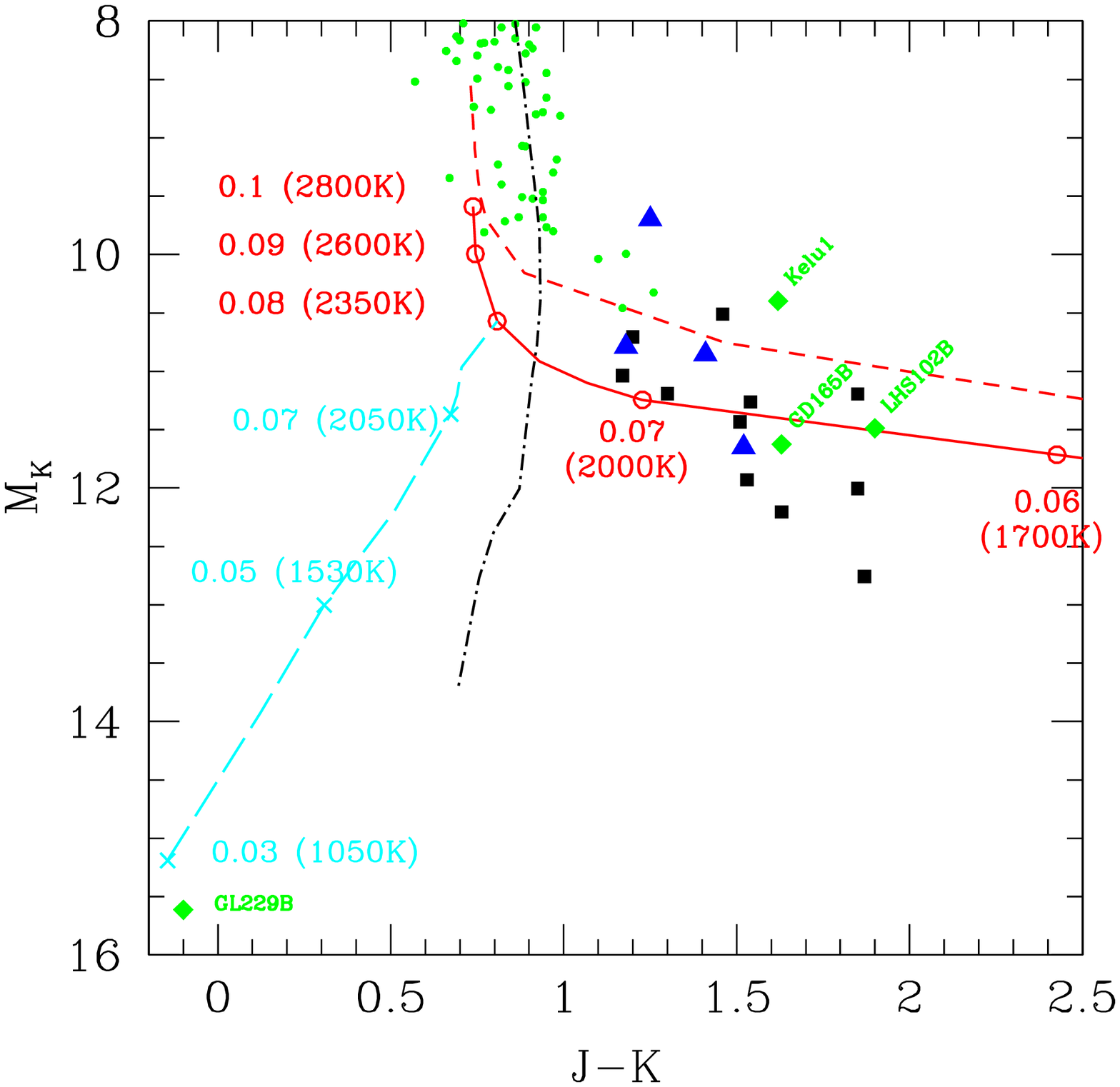}
\end{center}
\end{figure}

\vfill\eject
\begin{figure}
\begin{center}
\epsfxsize=180mm
\epsfysize=180mm
\epsfbox{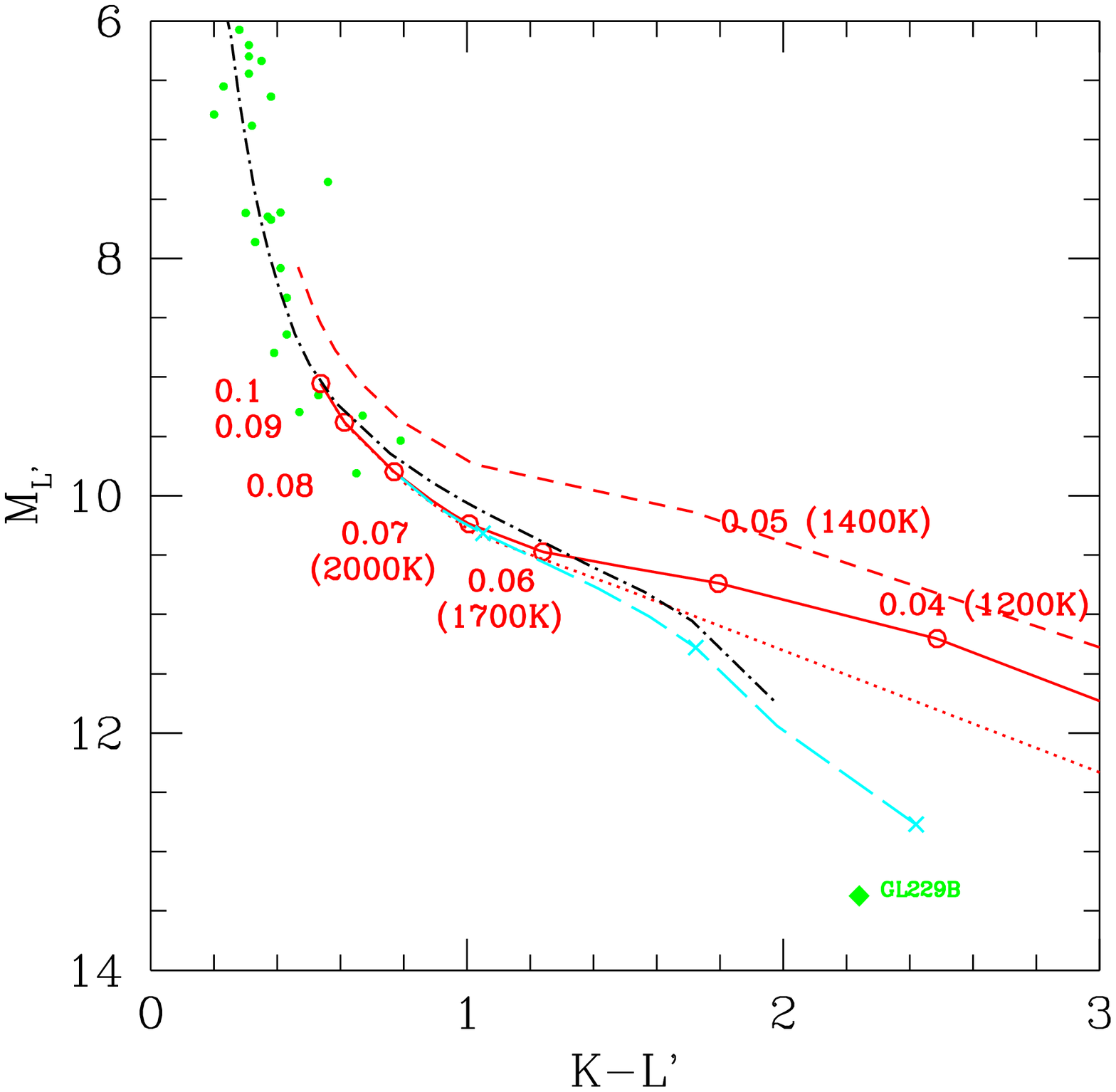}
\end{center}
\end{figure}

\end{document}